\documentstyle[11pt]{article}
  
\newcommand{\AmS}{{\protect\the\textfont2
  A\kern-.1667em\lower.5ex\hbox{M}\kern-.125emS}}

\title{{\bf
Ambiguities in QED:\\
Renormalons versus Triviality}}

\author{{{\sc D. Espriu}\thanks{espriu@greta.ecm.ub.es}
   \ and {\sc R. Tarrach}\thanks{tarrach@ecm.ub.es}}\\
       \llap{}%
       \small{ Departament d'Estructura i Constituents
               de la Mat\`eria}\\
       \small{ Universitat de Barcelona} \\
       \small{ and} \\
       \small{ Institut de F\'\i sica d'Altes Energies}\\
       \small{ Diagonal, 647}\\
       \small{ E-08028 Barcelona} }

\date{}

\begin{document}
\maketitle

\begin{abstract}

We point out that, contrary to what is believed to hold for
QCD, renormalons are genuine in QED; i.e. the ambiguities
which come with them do not require cancellation by
hypothetical non-perturbative contributions. They are just
the ambiguities characteristic of any trivial ---and thus
effective--- theory. If QED remained an isolated theory up to
an energy close to its triviality scale, these ambiguities
would surely hint at new physics. This not being so, the
renormalon ambiguities in QED lead to no new physics, not even
to non-perturbative contributions within QED itself.
\end{abstract}

% typeset front matter (including abstract)

\vfill
\vbox{
 UB-ECM-PF-96/8\hfil\null\par
 April 1996\hfil\null\par}

\clearpage

%%%%%%%%%%%%%%%%%%%%%%%%%%%%%%%%

Exactly as for $\lambda\phi^4$ there are two perturbative
quantum-field theoretical frameworks for dealing with QED.
If the approximation one uses is triviality insensitive,
such as finite-order perturbation theory, one can take the
continuum limit by removing the UV cutoff after having
renormalized the theory order by order in perturbation theory.
This procedure leads to quantitatively extremely successful
results, the paradigm being the lepton anomalous magnetic
moments. If the approximation used is triviality sensistive,
such as a summed-up infinite order perturbative computation, one
has to consider the theory as an effective one, with an UV
cutoff which cannot be removed. If one insists in removing it,
triviality makes the theory uninteresting.

In 1952 Dyson\cite{Dyson} gave a physically motivated
argument leading to the conclusion that in QED renormalized
perturbation theory has zero radius of convergence.
Nowadays one believes that the series is not only divergent, but
also non Borel-summable. As stressed by Parisi\cite{Parisi}
this pathology is related to the existence of the Landau
pole\cite{Landau}, or, in other words, to the fact that
perturbation theory cannot be trusted at large momenta. The
perturbative predictions are thus ambiguous. (To be sure
Dyson himself pointed out that due to the smallness of $\alpha$
these ambiguities would not prevent practical calculations
being made with the series, to an accuracy far beyond
current or even future experimental requirements.) These
ambiguities can also be seen to emerge as singularities of
the Borel transform of the perturbative series, called renormalons.
They have the structure of zero momentum insertions of local
operators\cite{Parisi}.

Renormalons were also introduced by 't Hooft to
further the understanding of QCD\cite{Hooft}. Unlike QED, where
renormalons are only of UV origin, there are both IR and UV
renormalons in QCD. The first ones lead to singularities, and
thus ambiguities, in perturbation theory, which call for
some non-perturbative contributions\cite{David}. (For instance,
for heavy quarks, the usefulness of the notion of a pole
mass\cite{Rolf} has been questioned by the unavoidable ambiguity of
order $\Lambda_{QCD}$ which goes with it\cite{Beneke}.)
The second ones lead to ambiguities in matrix elements
of local operators. It is now believed that renormalons
appear because the usual (and useful)
separation into coefficient functions
and matrix elements (short and long distance contributions,
respectively)
is not a clear cut one when regulators without a hard cut-off
are used.
When
a complete treatment is possible, IR and UV renormalons
eventually cancel, leaving an unambiguous prediction for
physical quantities\cite{Neubert}. These are good news for
QCD. Unfortunately, it is not so easy to pin down the
diagrams which in perturbation theory lead to singularities
in the Borel plane, let alone to use these families of diagrams
to somehow ``estimate" the natural size of non-perturbative
corrections. In practice one uses QED plus what
is called ``naive non-abelianization" to evaluate them; but
this approximation is somewhat suspect, as there is no known expansion
leading to such a simplification. Furthermore, one often
makes the logical leap that the  ambiguities in the
non-perturbative contributions are of the order of the
non-perturbative contributions themselves.

In QED everything is different, and this is what we want
to point out here. There is a well defined expansion
which leads to renormalons, namely the expansion in inverse powers
of the number of leptons.
The renormalon ambiguities
appear because one integrates over all momenta, being cavalier
with respect to the effective character of the theory (e.g.
by using dimensional regularization). If, on the contrary,
one treats the theory as an effective one, with an unremovable
UV cutoff, the ambiguities in summed-up renormalized perturbation
theory are the ones we expect in any effective theory. As we will
see in what follows, using this expansion
one can stay well within the framework of convergent perturbation
theory and so the issue of non-perturbative contributions which cancel
the ambiguities simply does not pose itself. Contrary to
QCD, the QED ambiguities are genuine but irrelevant. No
non-perturbative or indeed any physics can be learned from them.

An effective field theory with an unremovable dimensionful
UV cutoff $\Lambda$
may provide accurate predictions, but not infinitely
accurate ones\cite{Hooft2}, as some dependence on $\Lambda$
remains. We shall consider the large number of leptons limit of QED,
$N_l\to\infty$,
all of them with the same mass $m$ for simplicity. QED is then basically
solvable\cite{Coquereaux}. At the leading order in the expansion the sum
of all diagrams is equivalent to substituting the coupling constant by the
running coupling constant, which is given
by
\begin{equation}
\alpha(-q^2)={\alpha_0(\Lambda^2)\over{1+\Pi_0(-q^2,m^2,\Lambda^2)}}
\end{equation}
$\Pi_0$ is the usual one-loop photon bare self-energy. For
$\Lambda^2\gg-q^2\gg m^2$, eq. 1 reduces to
\begin{equation}
\alpha(-q^2)\simeq{\alpha_0(\Lambda^2)\over{1-{\alpha_0(\Lambda^2)\over
{3\pi}}\log{-q^2\over\Lambda^2}}}
\end{equation}
In the above expressions $\alpha_0(\Lambda^2)$ is the bare
coupling, or, if one wishes, the running coupling constant defined at
scale of the cutoff. The number of leptons has been absorbed in
the coupling, and is not explicitly written. The fine structure
constant is the zero energy limit of $\alpha(-q^2)$
\begin{equation}
\alpha\equiv\alpha(0)\simeq{\alpha_0(\Lambda^2)
\over{1-{\alpha_0(\Lambda^2)\over
{3\pi}}\log{m^2\over\Lambda^2} }}
\end{equation}
The running coupling constant in the large $N_l$ limit
has two interesting features. The first
one is triviality: $\alpha(-q^2)$ cannot be independent of
$\Lambda$ unless it is zero. Indeed, from eq. 2 it is clear
that, irrespective of $\alpha_0(\Lambda^2)$,
\begin{equation}
\lim_{\Lambda\to\infty} \alpha(-q^2)=0
\end{equation}
In other words, there is no strict scaling region for the
interacting theory. For theories for which the sign in the denominator
is the opposite there is no obstacle in taking
$\Lambda\to\infty$ while $\alpha(-q^2)$ is kept fixed. It is enough
to  ensure that
$\alpha_0(\Lambda^2)$ goes to zero adequately. This leads to a theory
which interacts even when the UV cutoff is removed, thanks to
asymptotic freedom. And it leads to the existence of
 renormalization group
invariants,  such as $\Lambda_{QCD}$.
On the contrary, there are no genuine
renormalization group invariants in large $N_l$ QED, except
those corresponding to a trivial free theory, such as the electron
mass.

There is however a limited scaling region. It is given by
$\Lambda<\Lambda_S$, where
\begin{equation}
\Lambda_S\simeq m\exp{{3\pi}\over{2\alpha}}
\end{equation}
This stems from requiring that eq. 3 is invertible, i.e. that
an $\alpha_0(\Lambda^2)$ exists such that $\alpha$ is
$\Lambda$-independent. Indeed for $\Lambda<\Lambda_S$
\begin{equation}
\alpha_0(\Lambda)\simeq{\alpha\over{1+{\alpha\over{3\pi}}\log{m^2\over
\Lambda^2}}}
\end{equation}
At $\Lambda=\Lambda_S$ $\alpha_0(\Lambda^2)$ diverges. This can
only be avoided by lowering $\alpha$ from its physical value
$\alpha=1/137$. This again leads to triviality, but in a way
in which one never uses the relation between $\alpha$ and $\alpha_0
(\Lambda^2)$ beyond the radius of convergence of the
corresponding perturbative series, while eq. 4 uses eq. 2 beyond
the radius of convergence. Therefore in the large $N_l$ limit
the relation given by eq. 6 is an exact one, free from
any hypothetical non-perturbative corrections. From eq. 3 one also
sees that massless QED ($m=0)$ implies $\alpha=0$,
but $\alpha(-q^2\neq 0)\neq 0$. In other words, in massless QED
the electron is completely screened at long distances.

As long as $\Lambda<\Lambda_S$ one can rewrite eq. 2
in the form of renormalized perturbation theory. For
$-q^2\gg m^2$
\begin{equation}
\alpha(-q^2)\simeq{\alpha\over{1-{\alpha\over{3\pi}}(\log{-q^2\over
m^2}-{5\over 3}+{\cal O}({m^2\over q^2}))}}
\end{equation}
Although this expression looks $\Lambda$-independent, it is,
for fixed $\alpha$, valid only if $\Lambda<\Lambda_S$.
At this point one should mention that the actual value
of $\Lambda_S$ is regulator dependent. Changing the
regularization
scheme may modify the constant pieces accompanying the logs in
eqs. 2, 3 (which we have not written up) and thus the relation
between $\Lambda_S$, $m$ and $\alpha$ given by eq. 5, but only
by subleading corrections.
Fixing unambiguously these
corrections would require
considering the relation between $\alpha$ and
$\alpha_0(\Lambda^2)$ at the next-to-leading order in the
large $N_l$ expansion. Such a relation is available in the
literature (see the third reference in \cite{Coquereaux}). It
should be mentioned that the next-to-leading beta function
in the large $N_l$ expansion
is given by a series in $\alpha$ with a finite radius of
convergence.

The second remarkable feature of the running coupling constant
is the existence of the Landau pole. The running coupling
constant blows up at high energies. From eq. 7 one finds
the Landau pole at $\sqrt{-q^2}=\Lambda_L\simeq\Lambda_S$.
Since the difference between $\Lambda_L$ and $\Lambda_S$
is regulator dependent,
we can take $\Lambda_L=\Lambda_S$ as the effective theory
is anyway quantitatively reliable only for $-q^2\ll \Lambda^2$ and the
regulator ambiguities are thus small.
For exactly the same reasons the effective theory will never
reach the Landau pole in its range of validity.

Let us now forgo the effective character of QED. Consider
an observable like the anomalous magnetic moment,
given in the large $N_l$ limit by\cite{Lautrup}
\begin{equation}
a={\alpha\over \pi}\int_0^1 dx {1-x\over{1-{\alpha\over \pi}
f(x) }}
\end{equation}
where
\begin{equation}
f(x)={1\over 3}[(1-{6\over x^2}+{4\over x^3})\log(1-x)-{5\over 3}
-{4\over x}+{4\over x^2}]
\end{equation}
with $-q^2/m^2=x^2/(1-x)$. Thus,
$-(\alpha/\pi)f(x)$ is the renormalized self-energy, and
\begin{equation}
{\alpha\over {1-{\alpha\over \pi}f(x)}}
\end{equation}
is nothing but $\alpha(-q^2)$. All divergences have been
subtracted in perturbation theory  and, indeed, at any finite order
in $\alpha$ $a$ is finite and unambiguous. (Up to terms of order higher
than those
actually evaluated. These are of course the usual perturbative ambiguities
related to the choice of a renormalization scheme.) However, after summing
up perturbation
theory, the integrand in the above
equation has obviously a singularity at the location
of the Landau pole
\begin{equation}
x_L=1-e^{-{3\pi\over \alpha}-{5\over 3}}
-4 e^{2(-{3\pi\over \alpha}-{5\over 3})}
+\ldots
\end{equation}
This singularity corresponds to the UV renormalon of
QED\cite{Lautrup2}. The prescription to handle this singularity
is of course ambiguous. For instance, following ref \cite{Lautrup2}
one could modify the integrand to
\begin{equation}
a={\alpha\over \pi}\int_0^1 dx {{x_L-x}\over{1-{\alpha\over \pi}
f(x) }}
\end{equation}
The modification is non-analytic and invisible in perturbation
theory, but cancels the singularity. But we could have taken another
prescription, such as the principal part. This would also lead to a well
defined ---but different--- result. The difference is of ${\cal
O}((1-x_L)^2)$. Do we learn
anything about the possible non-perturbative structure of the theory
from this? No.

In order to see this we have to remember that QED is
an effective theory. Then eq. 8 should be written as
\begin{equation}
a={\alpha\over \pi}\int_0^{1-k{m^2\over \Lambda_S^2}}
 dx {1-x\over{1-{\alpha\over \pi}
f(x) }},
\qquad k\gg 1,\qquad
\log k\ll{1\over \alpha}
\end{equation}
showing that the momentum is only integrated up to $-q^2
=\Lambda^2=\Lambda_S^2/k$, as befits an effective theory. Thus one
never integrates over the location of the Landau pole and
no ambiguities arise because of it. Furthermore, the
perturbative series is used within its radius of convergence
and hence there is no question of possible non-perturbative
contributions. However the result is of course not infinitely
accurate because the cutoff is arbitrary. The uncertainty
is of the order
\begin{equation}
\Delta a \simeq k^2 {m^4\over \Lambda_S^4}
\end{equation}
(modulo logarithmic corrections). If one pushes
the UV cutoff to its largest possible value, $\Lambda_S$,
the uncertainty becomes of the form
\begin{equation}
\Delta a\simeq \exp{-{6\pi\over \alpha}}
\end{equation}
exactly as the ambiguity due to the Landau pole.
The ambiguity disappears when $\Lambda\to\infty$, but then
the anomalous magnetic moment $a$ vanishes because
$\alpha\to 0$.

Of course the result of the previous equation is purely
perturbative and it is obtained without ever leaving the
region of convergence of the perturbative series in the
renormalized coupling. Because there are no non-perturbative
contributions, the previous result does not lead to
 any non-perturbative insight into the theory.
It is just a reflection of triviality.

Nothing prevents us from going to next order in the
large number of leptons expansion\cite{Smirnov}.
One encounters subdiagrams
which lead to the integral we have just been
discussing and the Landau singularity is still there.
These subdiagrams have also to be cutoff at
a scale $\Lambda^2$ as corresponds to the effective
character of QED (or, more specifically, of large $N_l$ QED).
The analysis goes through much as before.

A nice feature
of QED is that the leading and next-to-leading contributions
in the large number of leptons expansion can be worked out
explicitly. This is unlike in QCD where no known
systematic expansion displays the renormalon.
It has been
advocated that in QCD the analysis should mimick the one of QED
but replacing in an ad-hoc manner
$N_l/3$ by $\beta_1$, the first coefficient
of the beta function or even by
the beta function up to two loop order.The consequences
of such a daring approach have been analyzed recently\cite{Uraltsev}.
Nobody really knows whether such replacements are meaningful for
QCD.

In conclusion, the renormalon ambiguities of QED are nothing but
the ambiguities characteristic of a trivial, and thus
effective, theory. If QED remains an isolated theory,
they are there to stay. They will not be cancelled by any
other ambiguities. They do not contain non-perturbative
physics. They are just what characterizes triviality.
Furthermore in the specific case of QED they are even of
no use as harbingers of new physics, as at energies well
below $\Lambda_S$ (where any ambiguity is ridiculously small)
the theory is subsumed by the Weinberg Salam theory. This
theory has its own Landau poles, but this
is a different story. In the specific case
of the lepton anomalous magnetic moment,
a numerically  important part originates from the hadronic
contribution to the vacuum polarization $\Pi_0$, which has not been
considered either.

These results are not expected to be different for real QED, away
from the large number of leptons limit. They have been obtained within
a well defined expansion and within the radius of convergence
of perturbation theory. They also show that neither Landau
poles nor renormalons are problems in need of solving.

\bigskip\bigskip
We are thankful to J.L. Alonso, M. Asorey, J. Cort\'es, J. Esteve,
P. Hasenfratz, J.I. Latorre, A. Manohar, P. Pascual,
S. Peris and E. de Rafael for discussions.
We acknowledge the financial support of CICYT and CIRIT through grants
AEN95-0590 and GRQ93-1047, respectively.

\end{document}